\documentclass[useAMS,usenatbib]{mnras}
\usepackage{newtxtext,newtxmath}
\usepackage[]{hyperref}
\usepackage[dvipsnames]{xcolor}
\usepackage[utf8]{inputenc}
\usepackage{graphicx,url,times,subcaption}
\usepackage{cleveref}
\usepackage{orcidlink}
\usepackage[utf8]{inputenc}
\usepackage{anyfontsize}

\newif\ifpdf
  \ifx\pdfoutput\undefined
    \pdffalse
  \else
    \ifnum\pdfoutput=1
      \pdftrue
    \else
      \pdffalse
    \fi
  \fi

\ifpdf
    \graphicspath{{images/}}
\else
    \graphicspath{{eps/}}
\fi
\newcommand{\papertitle}{The life and times of dark matter haloes: what will I be when I grow up?
}
\newcommand{\shortpapertitle}{What will I be when I grow up?}

\ifpdf\hypersetup{
pdftitle={\shortpapertitle},
pdfauthor={Julian Onions},
pdfkeywords={N-body simulations, haloes evolution, dark matter},
pdfstartview=FitH,
}
\fi

\newcommand{\TheThreeHundred}{\textsc{The Three Hundred Project}}
\newcommand{\hkpc}{{\ifmmode{h^{-1}{\rm kpc}}\else{$h^{-1}$kpc}\fi}}
\newcommand{\hmpc}{{\ifmmode{h^{-1}{\rm Mpc}}\else{$h^{-1}$Mpc}\fi}}
\newcommand{\hgpc}{{\ifmmode{h^{-1}{\rm Gpc}}\else{$h^{-1}$Gpc}\fi}}
\newcommand{\Mvir}{{\ifmmode{\rm M_{vir}}\else{M_{vir}}\fi}}
\newcommand{\Msun}{{\ifmmode{\rm M_\odot}\else{M_\odot}\fi}}
\newcommand{\hmsun}{{\ifmmode{h^{-1} \Msun}\else{$h^{-1}$\Msun}\fi}}

\newcommand{\Fig}[1]{Figure~\ref{#1}}

\newcommand{\Sec}[1]{Section~\ref{#1}}

\title[\shortpapertitle]{\papertitle}
\author[Onions et al.]
{Julian~Onions\orcidlink{0000-0001-5192-6856},$^1$\thanks{E-mail:
    \href{mailto:julian.onions@gmail.com}{julian.onions@gmail.com}},
  Frazer~Pearce \orcidlink{0000-0002-2383-9250}$^1$, Alexander Knebe\orcidlink{0000-0003-4066-8307}$^{2,5,7}$, Meghan~Gray\orcidlink{0000-0002-6301-5870}$^1$, \newauthor Roan Haggar\orcidlink{0000-0001-5490-9621}$^{3,4}$, Ulrike Kuchner\orcidlink{0000-0002-0035-5202}$^1$, Ana Contreras-Santos\orcidlink{0000-0002-3374-4626}$^2$, Gustavo Yepes\orcidlink{0000-0001-5031-7936}$^{2,5}$,\newauthor Weiguang Cui\orcidlink{0000-0002-2113-4863}$^{2,5,6}$
\\
$^1$School of Physics \& Astronomy, University of Nottingham, Nottingham, NG7 2RD, UK\\
$^2$Departamento de Fısica Teórica, Módulo 15 Universidad Autónoma de Madrid, E-28049 Madrid, Spain\\
$^3$University of Waterloo. Department of Physics and Astronomy, University of Waterloo, Waterloo, ON N2L 3G1, Canada  \\
$^4$Waterloo Centre for Astrophysics, University of Waterloo, Waterloo, ON N2L 3G1, Canada\\
$^5$ Centro de Investigaci\'on Avanzada en F\'isica Fundamental (CIAFF), Universidad Aut\'onoma de Madrid, 28049 Madrid, Spain\\
$^6$ Institute for Astronomy, University of Edinburgh, Royal Observatory, Edinburgh EH9 3HJ, UK\\
$^{7}$International Centre for Radio Astronomy Research, University of Western Australia, 35 Stirling Highway, Crawley, Western Australia 6009, Australia\\
}

\begin{document}

\pagerange{\pageref{firstpage}--\pageref{lastpage}} \pubyear{2025}\volume{0000}

\maketitle

\label{firstpage}

\begin{abstract}
Are the most massive objects in the Universe today the direct descendants of the most massive objects at higher redshift? We address this question by tracing the evolutionary histories of haloes in the MultiDark Planck2 simulation. By following the 100 most massive halos at $z = 0$ across cosmic time, we find that only 40\% of them were among the largest 100 halos at $z = 1$. This suggests that many of today’s most massive clusters were not the most dominant structures at earlier times, while some of the most massive objects at high redshift do not remain in the top mass ranks at later epochs. The hierarchical nature of structure formation predicts that, on average, massive haloes grow over time, with their abundance in comoving space decreasing rapidly at higher redshifts. However, individual clusters exhibit diverse evolutionary paths: some undergo early rapid growth, while others experience steady accretion or significant merger-driven mass changes. A key assumption in self-similar models of cluster evolution is that the most massive objects maintain their rank in the mass hierarchy across cosmic time. In this work, we test this assumption by constructing a mass-complete sample of haloes within the $(1 \hgpc)^3$ volume of MultiDark and analysing when clusters enter and exit a high-mass-selected sample. Our results demonstrate that cluster selections must be carefully constructed, as significant numbers of objects can enter and leave the sample over time. These findings have important implications for observational cluster selection and comparisons between simulations and surveys, especially at high redshift.
\end{abstract}

\section{Introduction}

Galaxy clusters, as the largest gravitationally bound structures in the Universe, serve as key probes of large-scale structure and galaxy evolution in dynamic, complex environments. Their formation and growth are primarily governed by the surrounding dark matter density, with clusters assembling through the continuous accretion of material via cosmic web filaments and discrete mergers with smaller groups and clusters. This hierarchical buildup shapes their structural properties and influences the large scale distribution of dark matter haloes.

Theoretically, cluster formation follows a two-phase accretion model, where clusters undergo an initial fast accretion phase, dominated by major mergers, before transitioning to a slow accretion phase, characterized by the steady infall of diffuse material \citep{GunnGott72, Wechsler02, Zhao03}. In the early phase, rapid mass accumulation drives structural evolution, particularly within the cluster core. As accretion slows, the virial radius continues to grow, but the inner density profile stabilizes \citep{Ascasibar04, DiemerKravtsov14, Mostoghiu19}. This transition manifests as self-similar evolution, where virialized haloes maintain a roughly constant central structure over time.

\citet{Kaiser86} established this framework by proposing a one-dimensional mass homology, in which clusters evolve according to scale-free scaling relations. In this model, the characteristic cluster mass ($M_{halo}$) evolves as:
\begin{equation}
    M_{halo}(z) \propto (1+z)^{-\frac{6}{n+3}},
\end{equation}
where $n$ is the spectral index of the initial density fluctuations. This formalism predicts that higher-redshift clusters are denser, smaller, and dynamically hotter, consistent with X-ray and optical observations. The hierarchical nature of structure formation implies that massive clusters remain the most massive over time, with their comoving abundance decreasing rapidly at earlier epochs. While real clusters exhibit diverse evolutionary histories, this model underpins the assumption that the most massive objects at high redshift evolve into similarly massive structures today.

The statistical distribution of halo masses plays a fundamental role in understanding structure formation. \citet{PressSchechter74} introduced a pioneering formalism based on spherical collapse and Gaussian random field statistics, predicting a power-law behaviour at low masses and an exponential suppression at high masses due to the rarity of extreme peaks. However, this model underpredicts the abundance of massive clusters and overpredicts low-mass halos. \citet{ShethTormen99} refined this approach by incorporating ellipsoidal collapse, improving agreement with simulations, especially in the high-mass regime. \citet{Jenkins01} later demonstrated that the halo mass function is nearly universal when expressed in terms of the variance of the density field. These refinements have established mass functions as indispensable tools for predicting cluster abundances, strong lensing statistics, and large-scale structure growth.

Accurately measuring cluster masses remains a fundamental challenge in observational cosmology, as no direct observable provides an exact mass estimate. Various techniques infer masses based on galaxy dynamics, X-ray emission, gravitational lensing, and the Sunyaev-Zel'dovich (SZ) effect, each with distinct systematic uncertainties. Dynamical methods assume virial equilibrium but suffer from interloper contamination and anisotropic velocity distributions \citep{Wojtak07, Mamon13}. X-ray-based approaches, which assume hydrostatic equilibrium, tend to underestimate masses due to non-thermal pressure support \citep{Nagai07}. Weak lensing methods offer direct mass estimates but are affected by triaxiality and projection effects \citep{Mandelbaum06}. There are also modern approaches using machine learning to provide accurate and unbiased results such as those shown in \citet{Ntampaka2019,Yan2020, deAndres2022, Ferragamo2023}.

A promising alternative is the richness-based approach, which uses galaxy counts as a statistical proxy for mass. While effective, it requires precise calibration to account for environmental factors and variations in galaxy populations \citep{Old14}. Mass reconstruction uncertainties are lowest for richness-based techniques, though interloper contamination remains a concern, particularly in dynamical estimates. Mass uncertainties increase significantly for low-mass clusters (\( M \sim 10^{14} M_{\odot} \)), where the number of tracer galaxies is smaller, leading to mass scatter approaching an order of magnitude. 

Since dark matter dominates the mass budget of galaxy clusters, mass estimates rely on indirect proxies such as galaxy velocity dispersions, X-ray emission from hot intracluster gas, and weak gravitational lensing distortions. These proxies rank clusters by mass, establishing a one-parameter mass homology, which forms the foundation of theoretical scaling relations. However, the dynamical nature of cluster growth challenges this ranking, as mergers and accretion events redistribute mass, altering both the total cluster mass and its ranking within the homology framework. Observational limitations—such as large mass uncertainties and the snapshot nature of observations—further complicate comparisons with theoretical predictions.

Cluster catalogues are constructed using a variety of techniques, each targeting different physical properties of clusters. Optical and near-infrared surveys detect galaxy overdensities, often identifying clusters based on a prominent red sequence of early-type galaxies. These are combined with photometric or spectroscopic redshifts, as demonstrated in large-scale surveys such as SDSS \citep{Almeida_2023} and DES \citep{DES_overview}. X-ray surveys like those from ROSAT \citep{ROSATsurvey}, Chandra \citep{Kim_2006}, and XMM-Newton \citep{XmmGalaxy} trace the hot intracluster medium (ICM), while the Sunyaev-Zel'dovich (SZ) effect, measured in surveys such as ACT \citep{Act2016} and SPT \citep{SPTSZ}, provides a redshift-independent detection method. Weak lensing surveys (e.g., CFHTLenS, \citep{Shirasaki_2014}; DES lensing, \citep{Kilbinger_2013}) directly map the projected mass distribution through background galaxy distortions.

Each approach probes different cluster regions and has varying sensitivities to redshift and cluster size. X-ray surveys primarily trace dense central regions, whereas weak lensing captures the full gravitational potential. At high redshifts (\( z > 1 \)), galaxy densities decline as only the most massive galaxies remain detectable, leading to smoother overdensity fields and increasing difficulties in detecting clusters.

Theoretically, this different halo mass growth history can be characterized with one single parameter -- the halo formation time \citep[e.g.][and references thereafter]{Wechsler2002}. The halo formation time can be defined as the redshift when the halo accreted half of its final mass \citep[see e.g.][for different definitions and their relative differences]{Li2008}. Due to the different growth pace of the halos, their intrinsic properties also show differences; see the correlations with halo dynamical state \citep[e.g.][]{Mostoghiu19} and concentration \citep[e.g.][]{Wechsler2002}. Furthermore, the halo, as the host environment of galaxies, can influence the formation time difference and also influence the galaxies' properties, for example, the stellar mass of central galaxy \citep[e.g.][]{Matthee2017, Zehavi2018} and their colour \citep{Cui2021}, the ICL fraction \citep[e.g.][and references therein]{Kimmig2025} as well as the X-ray brightness \citep[e.g.][]{Cui2024}. We refer to (Srivastav et al. in prep.) for interested readers on quantifying the halo formation time with observables. The formation time difference is rooted in their large-scale environments \citep[e.g.][]{Tojeiro2017}, which in turn affect the galaxy clustering \citep[e.g.][and references thereafter]{Gao2005}, etc, known as assembly bias.

Given the issues and uncertainties in both the theoretical models and the issues of making accurate observations, 
simulations therefore play a crucial role in bridging theory and observations by evolving cosmological volumes over time. 
Large dark-matter-only simulations, such as MultiDark, provide (1 Gpc/$h$)\(^3\) volumes that capture the full cluster mass function. In this paper we track the most massive clusters from \( z=2 \) to \( z=0 \), examining their growth histories. While theoretical models predict global trends in cluster evolution, individual clusters follow diverse growth paths, some growing rapidly early on, while others exhibit steady accretion over time.

A key assumption in both observations and theory is that massive clusters masses can be predicted from early observations across cosmic time such as the example predictions in  \citet{Chiang_2014,Yuan_2014,Hatch_2016} and \citet{Toshikawa_2014}. This assumption is central to self-similar models of cluster evolution but has not been explicitly tested. In this work, we challenge this assumption by constructing a mass-complete sample of large clusters within the (1 Gpc/h)\(^3\) volume of the MDPL2 simulation, studying when clusters enter and exit a high-mass sample.

The paper is structured as follows: In \Sec{sec:data}, we describe the simulation setup and analysis methods. In \Sec{sec:results}, we present findings on cluster growth diversity. We discuss implications for mass-complete samples in \Sec{sec:realworld}, and conclude with a discussion of our results in \Sec{sec:conclusion}.

\section{Constructing the catalogues}\label{sec:data}

\subsection{The Data}

For this study we require a simulation with a large volume which allows a full range of clusters to be analyzed. High mass resolution is not needed because we are studying large clusters: what is required is sufficient resolution to probe a range of cluster masses. We choose the publicly available dark matter (DM) only MultiDarkPlanck2 simulation (MDPL2) \citep{klypin_MD}. MDPL2 is a periodic cube of co-moving length 1\hgpc\ 
containing $3840^3$ DM particles, each of mass $1.5 \times 10^{9} \hmsun$. The
Plummer equivalent softening of this simulation is $6.5 \hkpc$. The
cosmological parameters of the MDPL2 simulation are based on the
Planck 2015 cosmology \citep{planck2016}. 
The MDPL2 simulation has 50 snapshots regularly spaced between $z=2$ and $z=0$. For each of these snapshots, a halo catalogue has been extracted using Rockstar \citep{Behroozi_2013}. These halo catalogues are used to create a halo merger tree using ConsistentTrees \citep{Behroozi_2013b}, allowing the progenitors and descendants of every large halo to be identified.
Both the halo finding of Rockstar and the tree building of ConsistentTrees have been tested for reliability \citep{onions_2012,Srisawat_2013}. Throughout this work we use the most massive progenitor of any halo to define the main branch when tracing halos across cosmic time.

\subsection{Constructing the complete cluster sample}

\begin{figure}
  \includegraphics[width=\linewidth]{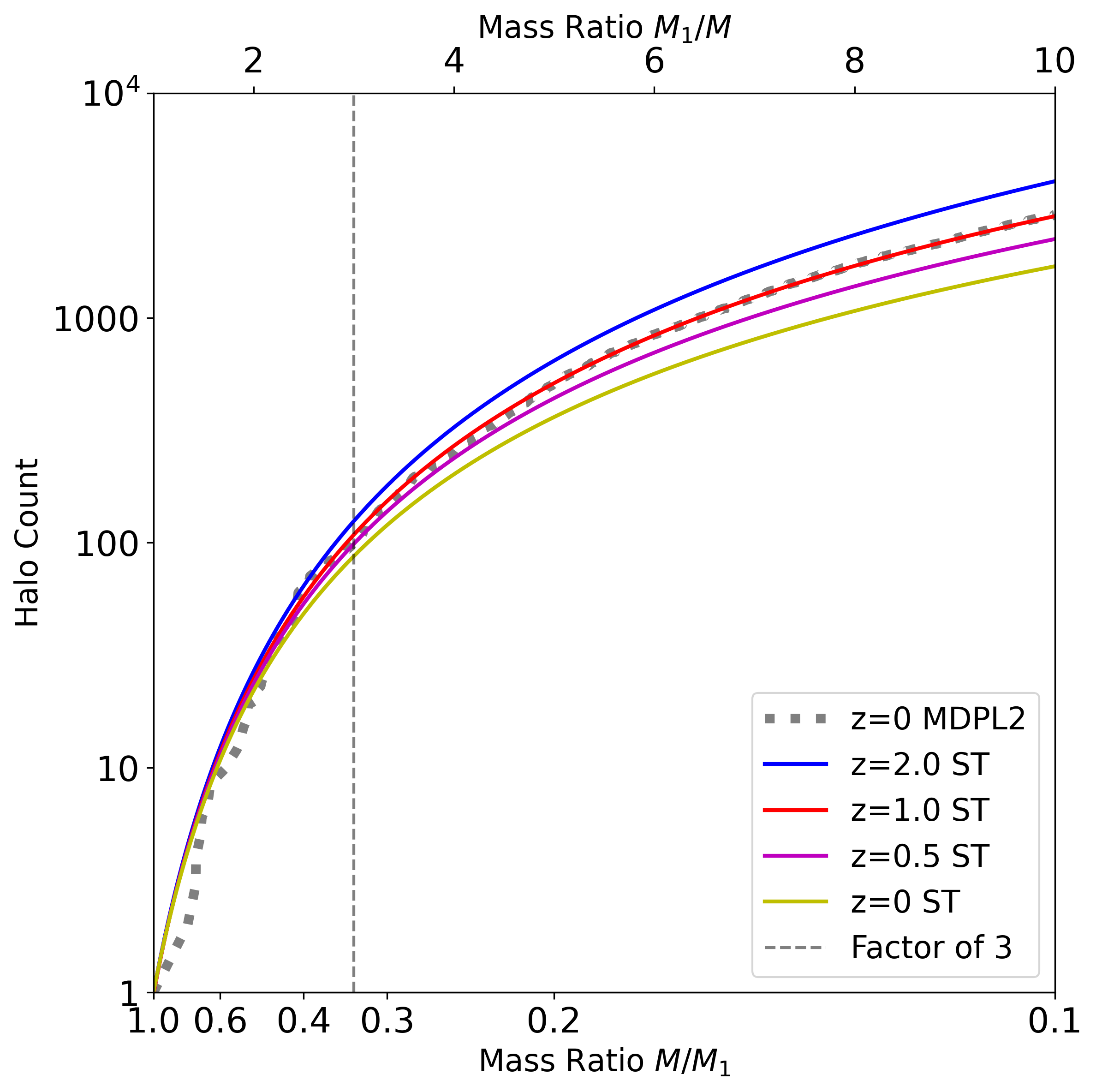}
  \caption{The cumulative number of halos expected within a volume of side 1Gpc/$h$ as a multiple of the mass of the largest halo expected as a function of redshift. The points indicate the measured mass ratio for haloes found within the MDPL2 volume at $z=0$. The curves indicate the analytic result found by integrating under the Sheth-Tormen mass function scaled to the MDPL2 volume (various redshifts as indicated). At low redshift $\sim 100$ halos are expected to lie within a factor of 3 in mass of the largest object in the MDPL2 volume.
  }
  \label{fig:STcounts}
\end{figure}

We begin by justifying which clusters we have selected as our `most massive' sample. \Fig{fig:STcounts} shows the cumulative number of haloes expected from the MDPL2 volume (1Gpc/$h$) as a multiple of the largest expected object  ($3.5 \times 10^{15} \hmsun$) compared to current object being considered. The dotted line shows the rank order of the current object for the halos found within the MDPL2 volume at $z=0$. 
As this is somewhat dependent on the mass of the currently most massive object we have also integrated under the analytic Sheth-Tormen mass function \citep{ShethTormen99}, scaled to the volume of the MDPL2 simulation (coloured curves at the redshifts indicated). 
The analytical result at high z is scaled to the most massive halo at that redshift. The Sheth-Tormen result is not entirely self-similar at higher mass ratios because at earlier times the mass function steepens slightly. 
Within the redshift range studied by this paper we expect to find around 100 haloes with masses within a factor of 3. This limit is a somewhat arbitrary number, but fits the curve reaching a more stable profile.\footnote{The lines plotted in \Fig{fig:STcounts} are scaled to one halo in the volume. This introduces sampling noise as the mass of the largest halo will vary from simulation to simulation as well as with the survey volume. The Rockstar catalogue extracts \citet{BandN} halos where the mass definition evolves with redshift. Exact agreement between the curves is therefore not expected.} 

We will therefore proceed by selecting the $N=100$ most massive clusters at all times for the purpose of this analysis. 
This band forms the 'froth' region, encompassing all the objects where one major merger can change the rank order of an object from relative obscurity to outstanding in a single event.

To create our {\it complete} cluster sample of the top 100 clusters across our redshift range, we need to identify all clusters that were ever within this froth region encompassing the largest 100 clusters. To achieve this, we use the halo catalogs from earlier snapshots sorted by mass. In this way, the identities of the largest clusters at all times can be found. Many of these objects are the same physical cluster seen at different times. We can identify these correspondences using a merger tree which connects objects at different times using the list of particles they contain. 

\section{Churn in the cluster sample}\label{sec:results}

\begin{figure}
  \includegraphics[width=\linewidth]{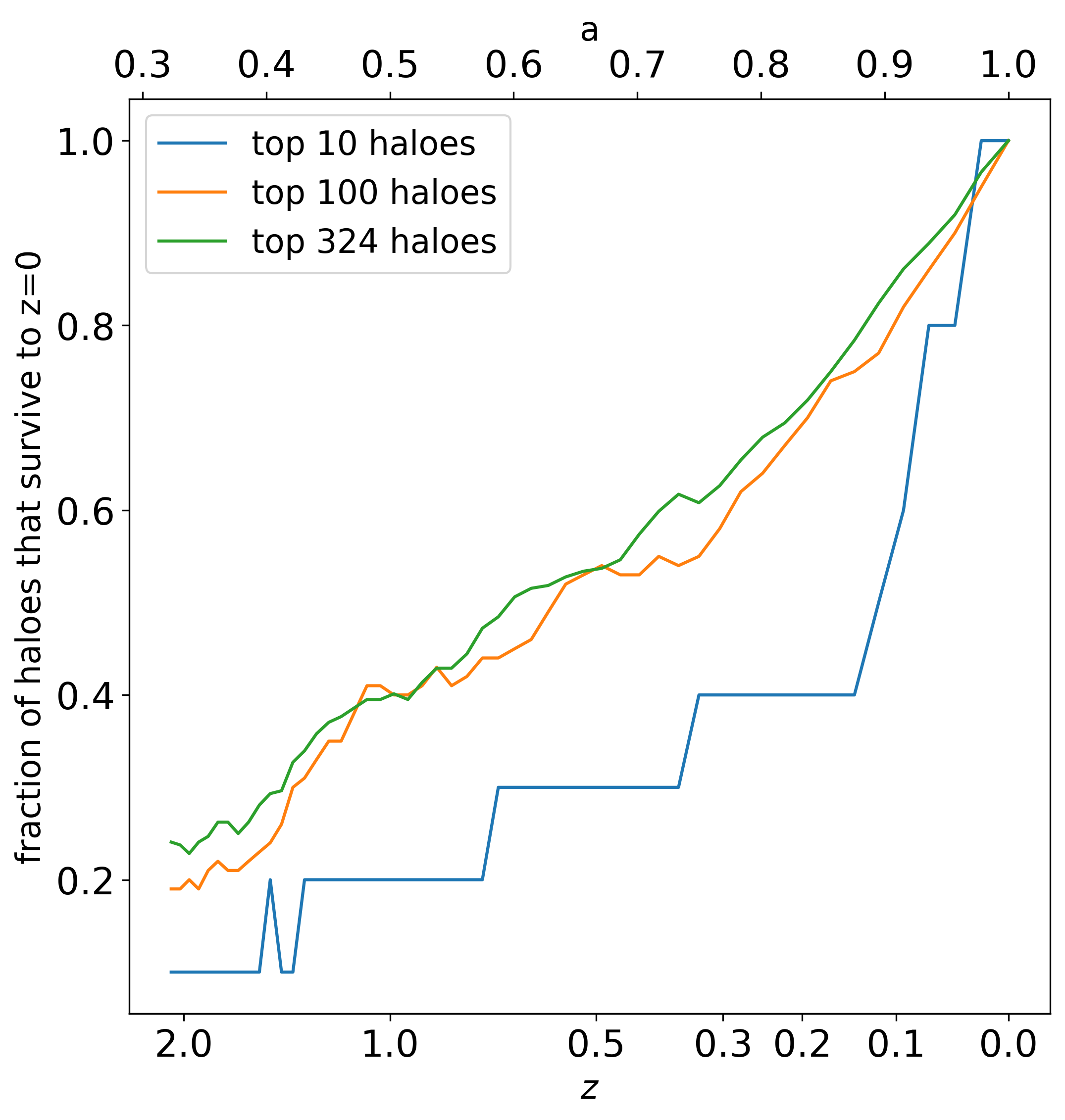}
  \caption{The fraction of the haloes in the top $N$ at $z=0$ that remain so at higher redshifts. We compare the rates for the top 10, 100 and 324 haloes at each redshift. By $z=1$ only $\sim$40\% of the 100 largest clusters at $z=0$ remain in the sample.  
  }
  \label{fig:progenitor}
\end{figure}
\begin{figure}
  \includegraphics[width=\linewidth]{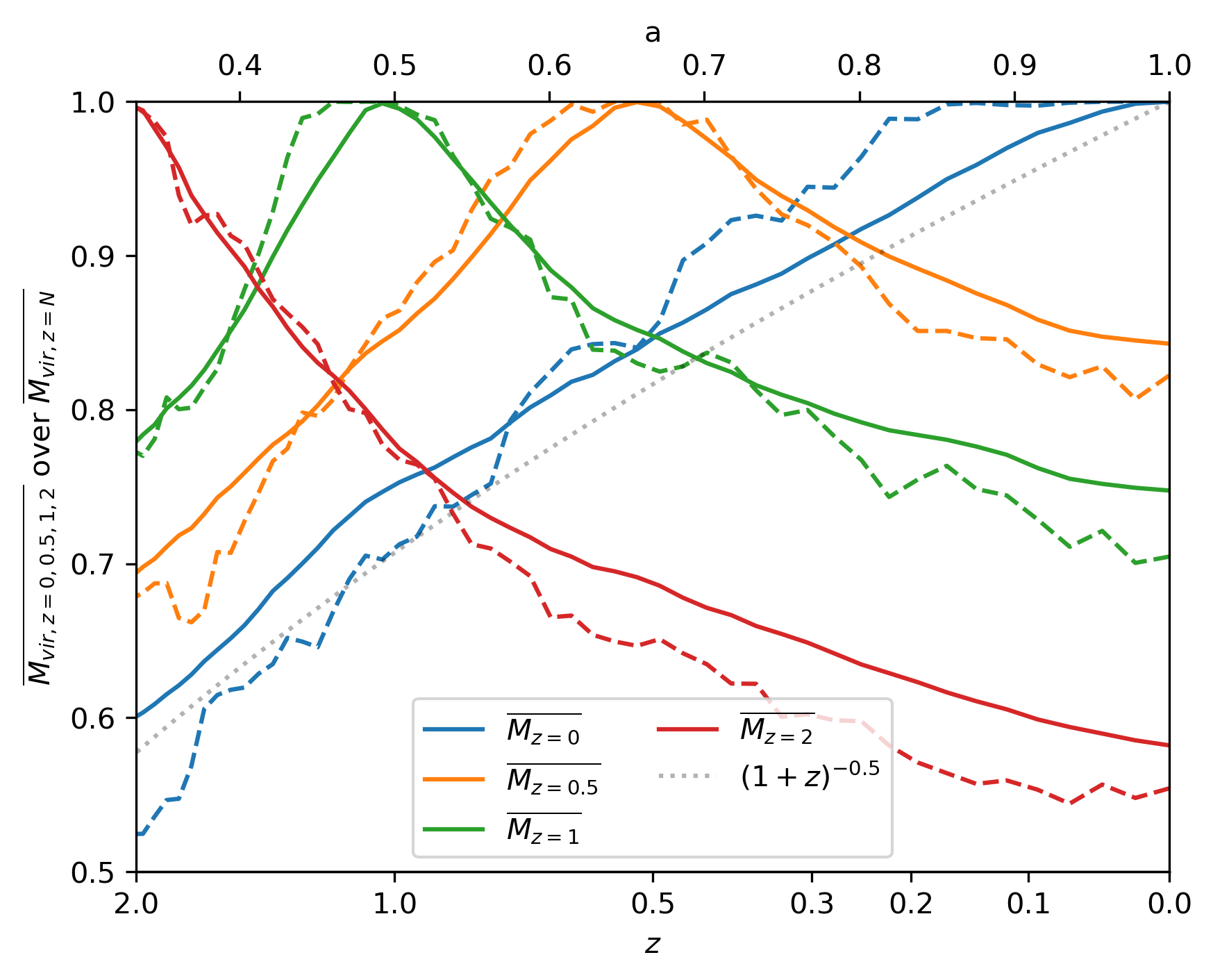}\hfill    
  \caption{Mass evolution of the top 100 clusters at different epochs from $z=2$ forwards as a fraction of the mass of the largest 100 haloes at the indicated redshift. The blue, orange, green, and red lines show the fractional mass of a fixed sample of the largest 100 haloes selected at $z=0,0.5,1,2$ respectively as they evolve over time, relative to the mass of the top 100 clusters selected at any given redshift $z$. Solid lines show mean mass, dashed line show median mass. The dotted grey line shows the curve for $(1+z)^{-0.5}$.}
  \label{fig:RSevolve}
\end{figure}

\begin{figure*}
 \includegraphics[width=0.45\linewidth]{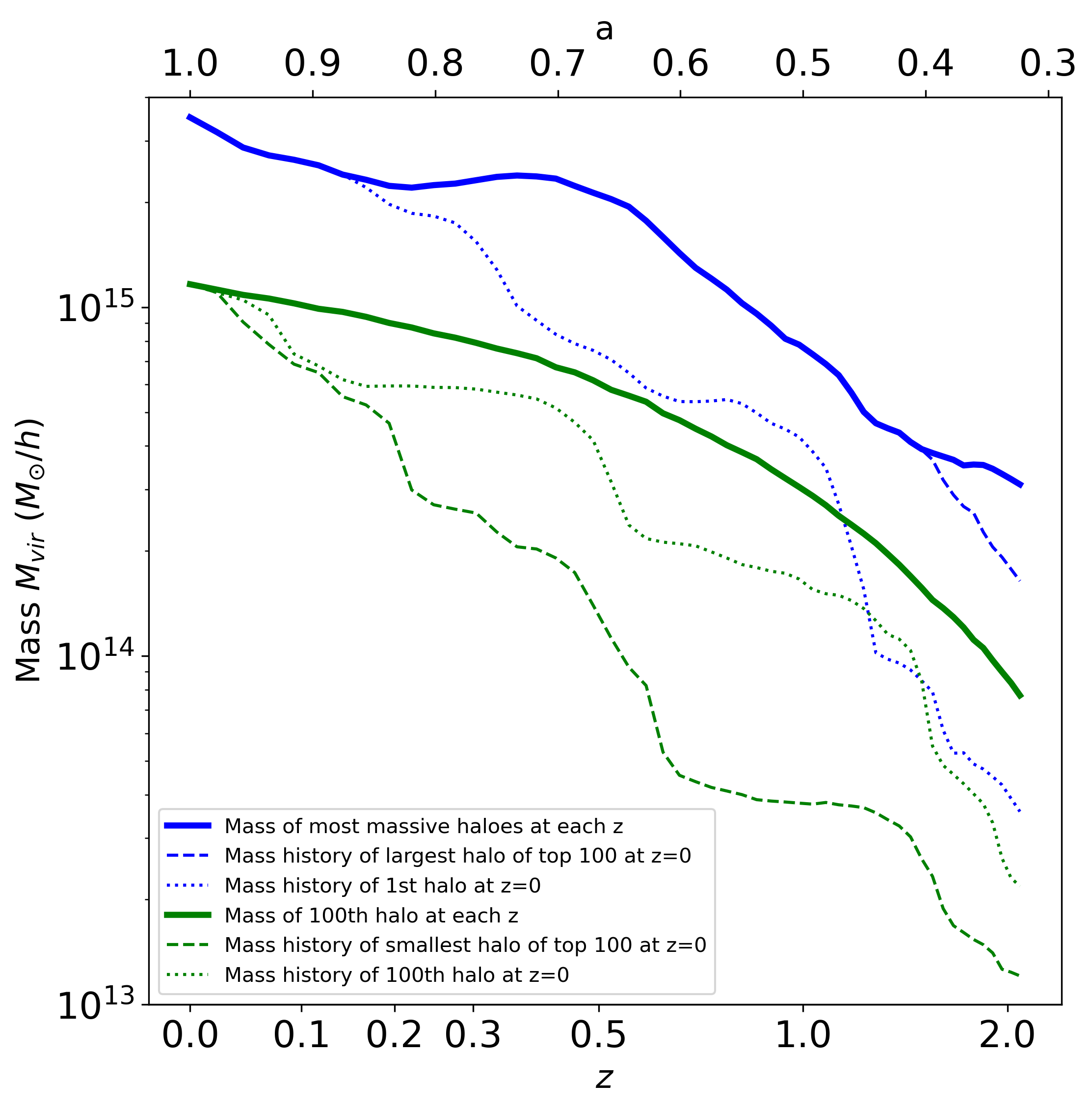}
 \includegraphics[width=0.45\linewidth]{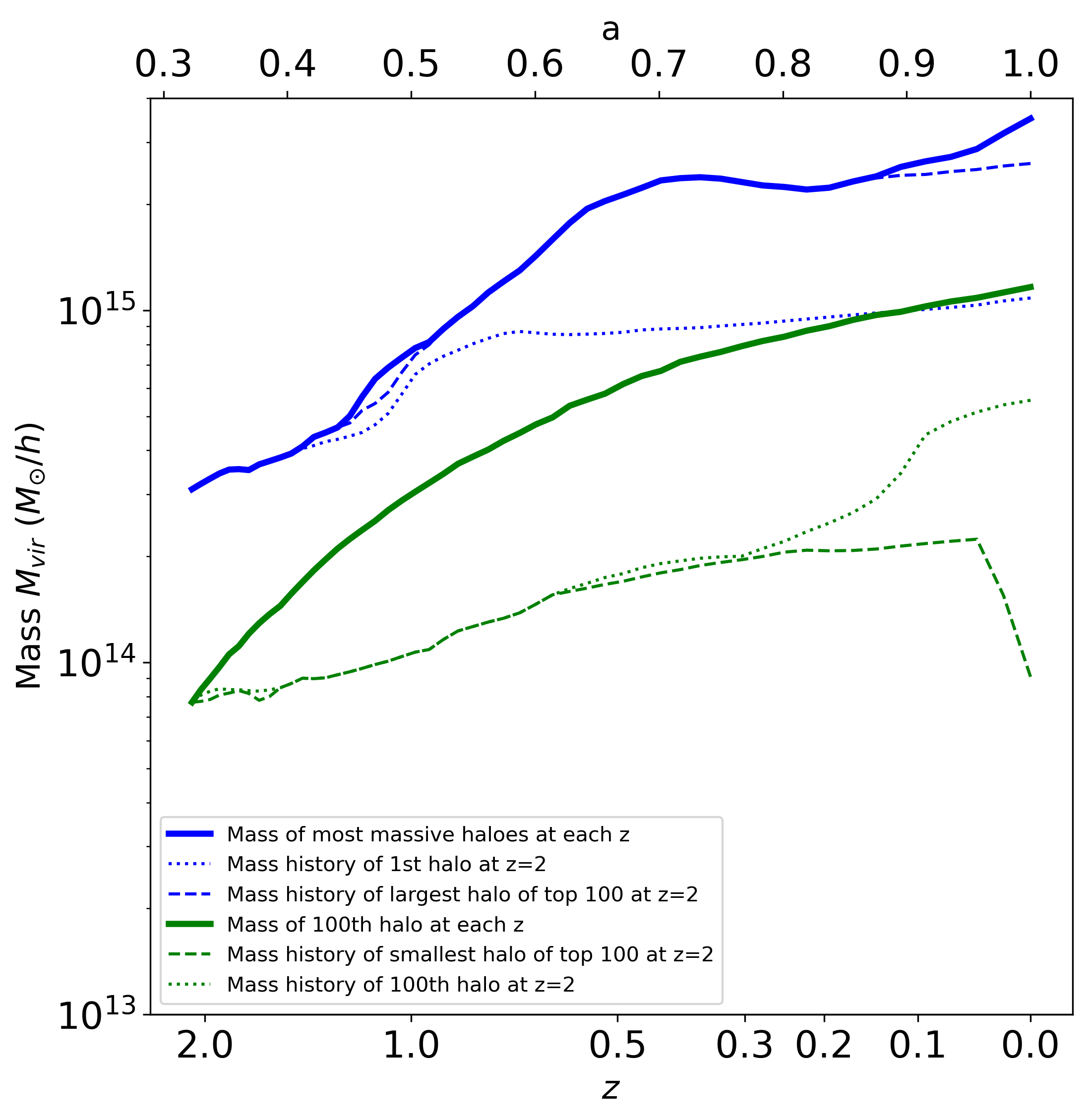}
  \caption{The mass evolution of various quantities. The left panel is for samples selected at $z=0$ tracked back to $z=2$. The right panel is the reverse: i.e. samples selected at $z=2$ tracked to $z=0$. 
  The various samples are: blue solid line: the most massive object at the indicated time. 
  Blue dashed line: maximum mass of any member in the original sample at this time. 
  When this differs from the blue line, none of the original sample are currently the largest object at that epoch. 
  Blue dotted line: the mass history of the largest halo at chosen epoch ($z=0$ or $z=2$).
  Green solid line: mass of the 100th most massive halo at the indicated time. 
  Green dashed line: the minimum mass of the top 100 haloes at each time. 
  Green dotted line: mass history of the 100th smallest halo at the chosen epoch ($z=0$ or $z=2$). 
  }
  \label{fig:masslimit}
\end{figure*}

\begin{figure}
  \includegraphics[width=\linewidth]{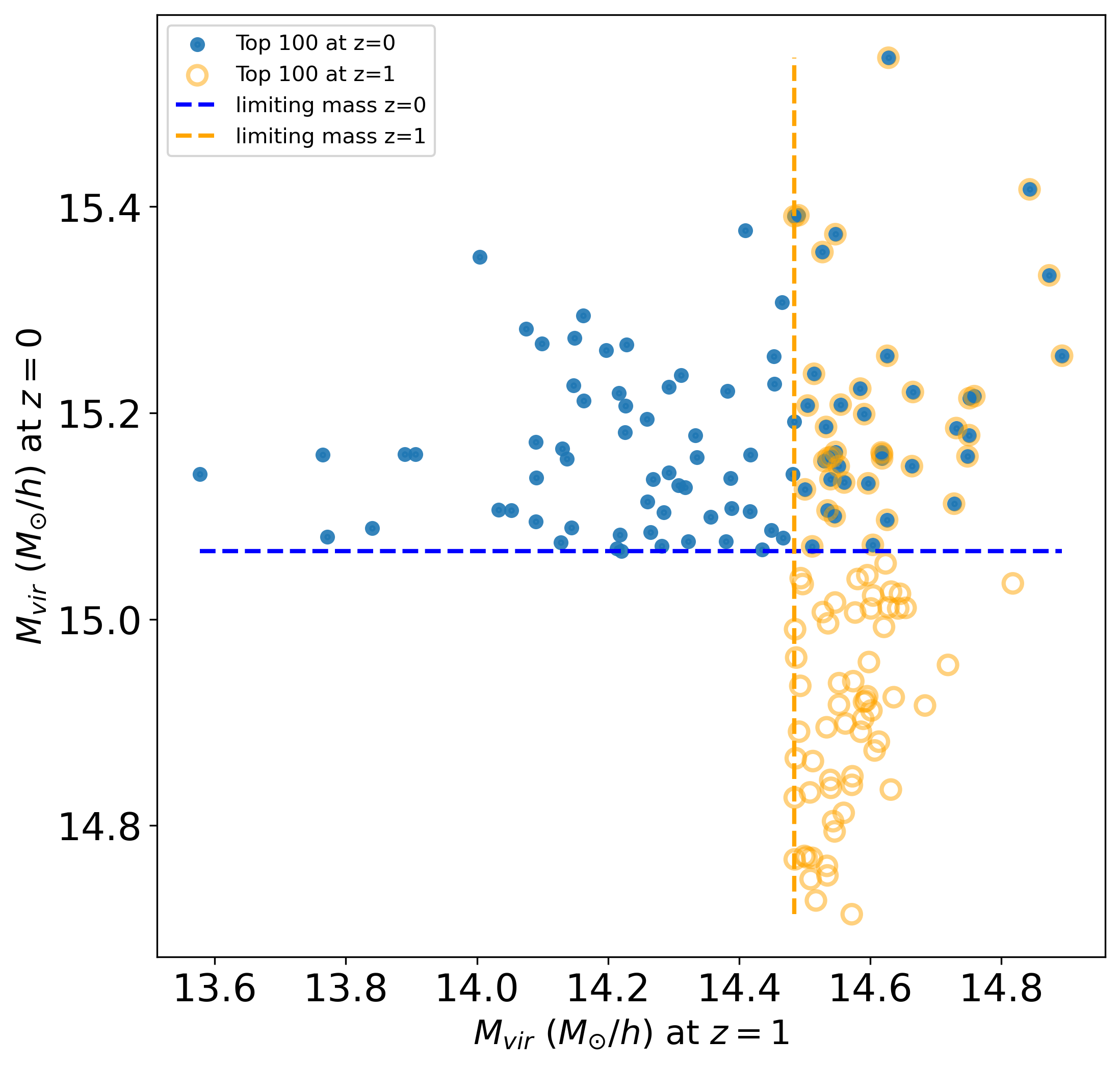}
  \caption{Mass at $z=0$ compared to mass at $z=1$ for the largest 100 haloes. Smaller blue filled circles indicate objects within the largest 100 at $z=0$ and their progenitor mass at $z=1$. Larger open orange circles indicate the objects found within the top 100 objects at $z=1$ and their descendent mass at $z=0$. Horizontal and vertical dashed lines indicate the limiting mass of the respective samples.}
  \label{fig:massz0z1z2lines}
\end{figure}

By tracing back the evolution of the most massive haloes at redshift $z=0$, we find that the assumption that the most massive objects always remain so is clearly incorrect. The fraction of objects that are still in the top $N$ objects as a function of redshift is shown in \Fig{fig:progenitor}. It is apparent from this that the most massive progenitors do not always become the most massive objects at the present day and that the number that do survive reduces significantly as we look further back in time. This is the `churn' of clusters that move in and out of the selection as mergers occur. For $N=100$ we see that $40\%$ of the haloes have fallen out of the sample by $z=1$. Although noisy, there is even more significant churn amongst the $N=10$ most massive objects where 60\% have dropped out of the sample by $z=0.2$. There is little difference in the dropout rate between $N=100$ objects and $N=324$ objects, which is the number of clusters used in \TheThreeHundred\ \citep{the300}. 

In \Fig{fig:RSevolve} we compare the mass evolution of various fixed cluster selections defined at different epochs, to the mass of the largest 100 clusters at the selected time. Specifically, we select fixed cluster samples of the most massive 100 clusters at $z=0, 0.5, 1$ \& 2, indicated by the blue, orange, green and red lines respectively, and then trace their mass evolution forward and backwards in time. Solid lines indicate mean mass (the mean mass of the selected haloes in each redshift snapshot), dashed lines indicate median mass. Each line rises to match the mass of the largest 100 clusters at the selection epoch as expected. Either side of this time the mass of the fixed sample falls away. Thus the mass of all of the fixed selections (the top 100 clusters at a fixed redshift) is less than that of the evolving selection (the top 100 clusters at any given redshift), and therefore growing more rapidly than the evolving selection prior to the selection epoch and more slowly post the selection epoch. For mean mass, the rates of growth of the fixed samples are more rapid than the evolving sample prior to the selection epoch by a factor of $(1+z)^{-0.5}$. With reference to the theoretical work of \citep{Kaiser86} (Equation~1 above) we see that this is a significant difference in mass growth rate between the samples. At $z=1$ the fixed sample defined at $z=0$ has only $\sim 75\%$ of the mean mass of a similarly sized sample of the largest objects in the volume.
Note - the mass definition used here and later is the Rockstar virial mass, $M_{vir}$, which is derived from the \citet{BandN} definition. 

The evolution of the mass for various samples, both backwards in time from $z=0$ (left panel) and forwards in time from $z=2$ (right panel) are shown in Figure~\ref{fig:masslimit}. The top blue line, which is the same in both panels, tracks the maximum mass of any halo in the volume at the given time. The dashed blue line, which is for the most part underneath the blue line, indicates the maximum mass of any member of the original sample (at $z=0$, left panel, and $z=2$, right panel). When this dashed blue line is visible, this indicates that the largest object in the volume was not part of the original sample. For the $z=0$ sample, this only occurs at redshifts above 1. Conversely, for the $z=2$ sample, none of the 100 largest objects at this time evolve into the largest object below $z=0.1$. 
The dotted blue line tracks the evolution of a single object, the most massive in the original sample. For the $z=0$ sample, the largest object loses this status around $z=0.15$ and gradually falls away. At $z=2$ it has a mass close to $1/10^{th}$ of the largest object at that time. In the right hand panel, the most massive object at $z=2$ stays near the top of the mass ranking until $z\sim 1$ and then experiences a prolonged period of little mass growth, ending up with a mass less than a third of the largest object at $z=0$. 

A second set of curves start from a lower point, marking the minimum mass of the 100 objects that make up each sample. The green solid line indicates the minimum mass of any object in the largest 100 at the given time. This line is the same in both panels. The mass limit of the sample drops from $\sim 10^{15}\hmsun$ at $z=0$ to $8 \times 10^{13}\hmsun$ at $z=2$. 
The final two curves mark, in dotted green, the mass of the smallest object in the original sample. Note that there is a period of time around $z=1.5$ in the left panel when the smallest object in the sample at $z=0$ is actually more massive than the progenitor of the largest object at $z=0$ (green and blue dotted lines cross). 
The green dashed lines show the mass evolution of the smallest of any of the objects that form the original sample. For example, one of the haloes that is in the top 100 at $z=0$ has a mass close to $10^{13}\hmsun$ at $z=2$. 
Conversely, the green dashed line in the right panel shows a rapid decrease close to $z=0$. This is an example of a halo becoming rapidly stripped as it merges with an even larger halo. 

We directly compare the masses of the top 100 objects at $z=0$ to the mass of the main progenitor of these clusters at $z=1$ (blue filled circles) in \Fig{fig:massz0z1z2lines}. Open orange circles indicate the masses of the top 100 objects at $z=1$ and their descendent mass at $z=0$. Objects that appear in both lists are then readily distinguishable, being marked by blue points with an orange surround. It is evident that while, as we saw in \Fig{fig:masslimit}, essentially all the objects grow in mass as time progresses the rate of growth is very different on an object-by-object basis. The limiting mass of both samples is indicated with the horizontal and vertical dashed lines. Even for objects that appear on both lists there is significant shuffling in the rank order. 

\subsection{Prevalence of major mergers and frequency of unrelaxed clusters}

\begin{figure*}
\begin{subfigure}{0.45\linewidth}
  \includegraphics[width=\linewidth]{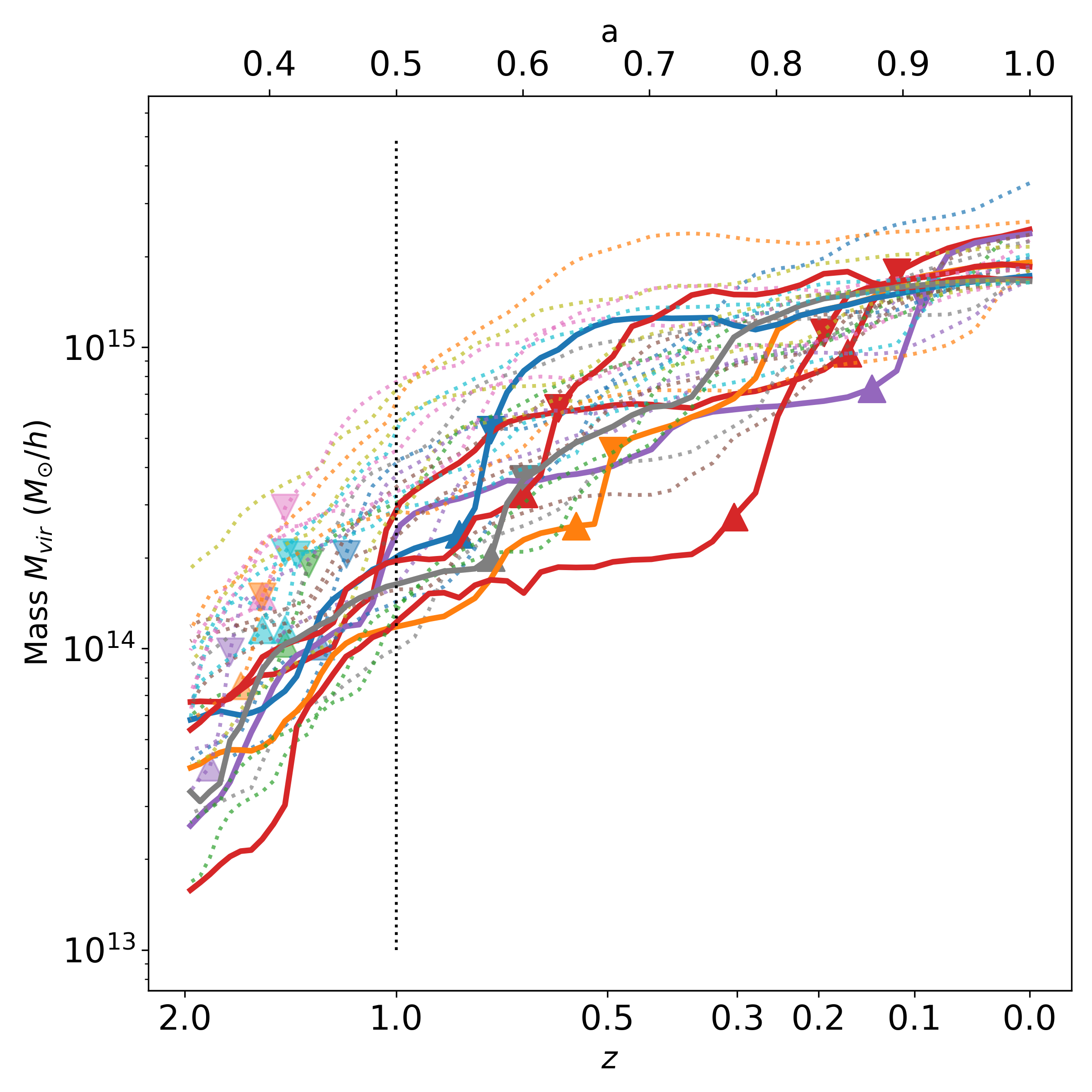}
    \caption{The top 30 haloes at z=0 mass growth}
    \label{fig:haloMM1}
\end{subfigure}
\begin{subfigure}{0.45\linewidth}
    \includegraphics[width=\linewidth]{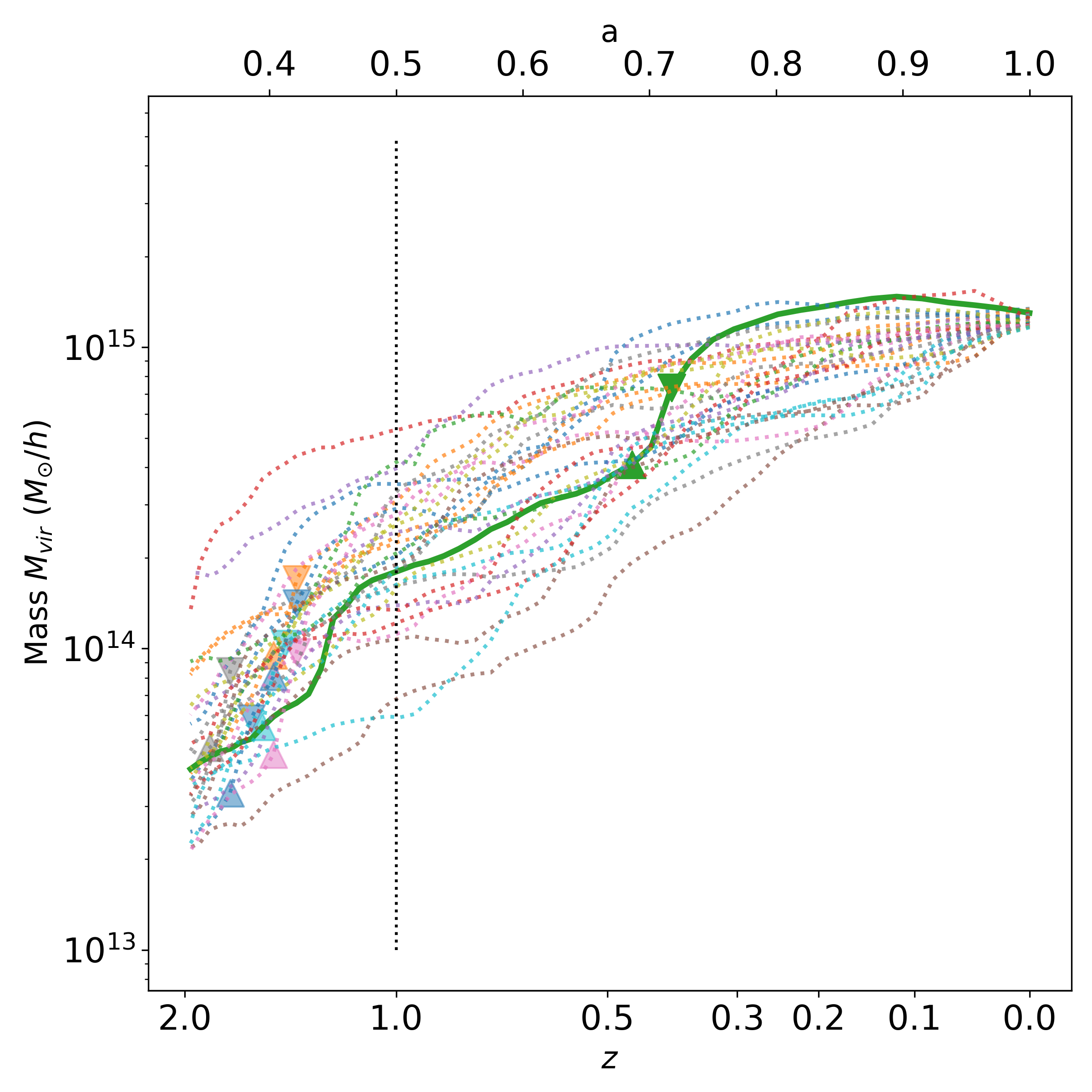}\hfill
    \caption{The top 70-100 haloes at z=0 mass growth.}
    \label{fig:halomm2}
\end{subfigure}
    \caption{The mass growth of haloes. Left panel (a): the largest 30 haloes in the volume at $z=0$. Clusters undergoing a major merger since $z=1$ (shown via the dotted vertical line) are indicated as solid curves with the start and finish time of the merger indicated via the filled triangles. Dashed curves indicate clusters without a major merger in this time interval. Major mergers are defined via a fractional mass increase technique described in \citet{Contreras_Santos_2022}. The right panel (b) gives mass growth histories for clusters between 70 and 100 in mass rank order at $z=0$. As with the left panel, the bold lines indicate clusters with a major merger since $z=1$. Only one object has experienced a major merger, as opposed to 7 of the first 30 haloes.}
    \label{fig:HaloMerger}
\end{figure*}

\begin{figure}
  \includegraphics[width=\linewidth]{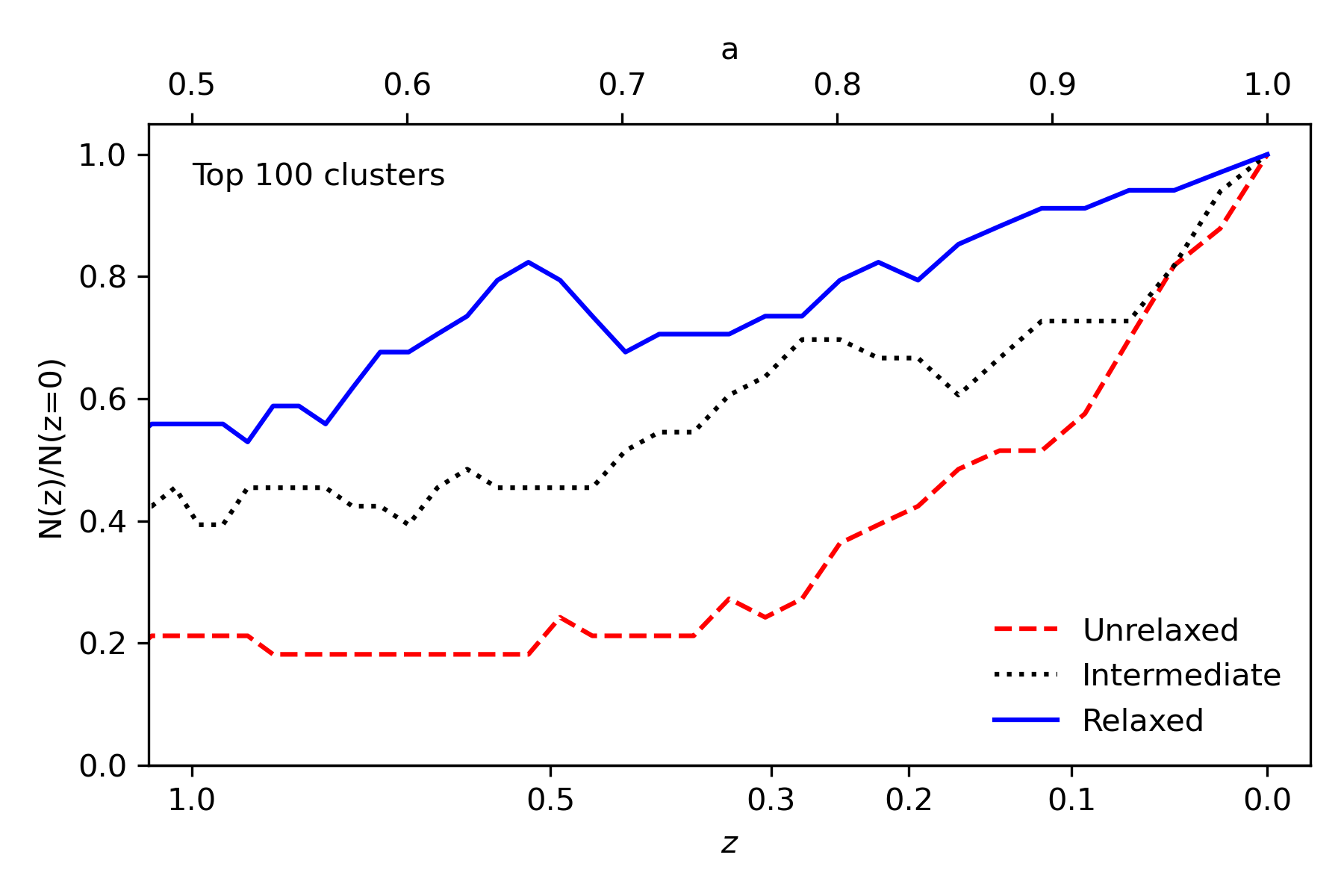}
\caption{Splitting the 100 clusters in the $z=0$ sample into three roughly equal sized bins according to dynamical state (taken from \citet{haggarDS}), the fraction of each of these groups that remain in the top 100 sample at earlier times. Most of the unrelaxed clusters enter the sample at late times whereas the relaxed group enters the sample much earlier on average.}
\label{fig:dstate}
\end{figure}

We can use the mass accretion history of each object in our $z=0$ sample to check the contention that objects near the very top of the mass function have an enhanced likelihood of having recently experienced a major merger. In order to test this, we track the mass growth of the very largest haloes (left panel) and a similarly sized sample from slightly further down the mass function (right panel) in \Fig{fig:HaloMerger}. We locate the major mergers experienced by these haloes using the fractional mass growth technique described by
\citet{Contreras_Santos_2022}, marking the onset and finish of each event with filled triangles. Mass growth from $z=1$ histories containing major merger events are shown as solid lines, those without are shown as dotted lines. The very largest clusters clearly experience more major merger events than the clusters from further down the mass function. Amongst the collection of clusters present in the volume, these major merger events raise individual objects to the very top of the ranking order. We therefore expect a raised fraction of the highest ranked objects to be dynamically disturbed systems.

We can further test this by examining the measured dynamical state of each of the 100 clusters in the $z=0$ sample (using equation one taken from \citet{haggarDS}). We correlate this dynamical state measure with the time each cluster became one of the largest 100 clusters in the MDPL2 volume. By splitting the sample via relaxedness into 3 sets (containing 33,33 \& 34 objects) we see in \Fig{fig:dstate} that the clusters that are identified as being the most unrelaxed at $z=0$ enter the sample much later than those clusters that are classified as being the most relaxed at $z=0$. This is what we expected: clusters that enter the sample late are likely to be undergoing a recent major merger.

\section{The practicalities of building a sample that is mass-complete at all redshifts}
 \label{sec:realworld}
As we have shown above, a cluster sample selected at $z=0$ will not contain all of the largest haloes at earlier times. This limits the capacity to study the full dynamic range of cluster assembly histories, as well as not optimally matching samples selected observationally at higher redshifts. In this section we discuss the practicalities of constructing an extended sample of clusters that is mass complete at {\em all} redshifts back to $z=1$. We do this within the context of \TheThreeHundred.

\subsection{Fixing the redshift range \texorpdfstring{$0<z<1$}{0<z<1}}

Our choice of a limit of $z=1$ is motivated by the recognition that at earlier times, selecting a single massive collapsed object is not a successful approach to characterising large-scale overdensities, or protoclusters. If we were to increase the limit to $z=2$, we observe that only 29 haloes would be in common with the top 100 at $z=0$, or 84 in the top 324.  Furthermore, the smallest cluster from the top 100 selected at $z=2$ is inconsequential by $z=0$, ranking $31,760^{th}$: this is a result of mass stripping at late times. The next smallest evolves into the $\sim\!5,000^{th}$ largest halo at $z=0$.

Due to the dramatic evolution of individual halo masses over this redshift range, observational selections based on galaxy overdensities at epochs $z>1$, would therefore typically involve smoothing the observed galaxy density over much larger volumes. Such a process could be implemented within our dataset to explore the progenitors of massive clusters back to $z=2$. We reserve that discussion for a future paper and focus here on the construction of cluster samples that satisfy the condition of being complete at all redshifts between $0<z<1$, considering (1) a redshift-varying mass-completeness limit; and (2) a fixed mass completeness limit.

\subsection{A sample with an evolving mass limit: the \texorpdfstring{$N$}{N}-most massive clusters at each redshift}

Having fixed the redshift range across which the sample will be complete, we now consider the effects of defining the sample by fixed number, namely $N$=100. The solid green line in \Fig{fig:masslimit} illustrates the evolving mass completeness limit when the sample of clusters selected from the MDPL2 parent volume is restricted to the top 100 at each redshift. This limit ranges from $10^{14.5}\hmsun$ at $z=1$ to $10^{15.05}\hmsun$ at $z=0$. In practical terms, to include all clusters that satisfy this selection we would need to include a total of 268 regions. Of these regions, only 68 are not already contained within our existing resimulations of the 324 largest haloes at $z=0$ from \TheThreeHundred. 

If we consider more closely the demographics of the resulting superset of 268 clusters, half of the largest 100 clusters selected at $z=0.5$ remain in the top 100 at $z=0$, while 87 remain in the top 324 largest clusters at $z=0$. The smallest cluster in this sample selected at $z=0.5$ falls from $100^{th}$ largest to $688^{th}$ largest when evolved to $z=0$, far below the evolving mass completeness limit at that redshift. From $z=1$, 42 clusters remain in the top 100 at $z=0$, with 71 in the top 324. The smallest in the sample at $z=1$ becomes the $1,105^{th}$ largest by $z=0$. Therefore, the superset of 268 clusters contains not only the top 100 clusters at $z=0$, but also comprises a considerable number of clusters that were once massive relative to their peers but then experienced slower late-time mass accretion histories leading to more moderate final masses at $z=0$.

\subsection{A sample with a fixed mass limit}

However, neither the original fixed sample of clusters selected at $z=0$ nor the complete sample of all clusters that were {\em ever} one of the top 100 largest objects at any redshift perfectly mimics an observational selection strategy. Such a sample would ideally impose, for example, a fixed minimum mass for inclusion \citep[e.g.][]{Vikhlinin_2009}. While the planned extension to \TheThreeHundred\  sample will proceed as described in the previous subsection, it is worth considering the practicalities of imposing a single mass limit covering all redshifts. 

As \Fig{fig:masslimit} shows, the minimum mass required for inclusion in the complete sample varies significantly with redshift. Even the largest cluster at $z=1$, at $10^{14.88}\hmsun$, would fail to meet the mass threshold of the 100 most massive object at $z=0$ $(10^{15.05}\hmsun)$. Few objects enter a mass-limited sample and then fall out again: such an event would require those haloes to not grow as fast as others. While this may indeed occur (cluster masses are naturally overestimated by the halo finder at the peak of a merger by a factor of around 50\% due to the (by definition) evolving nature of the merger process) the general trend is for a monotonic increase in mass as time progresses. The churn in the membership of our cluster samples is primarily caused by the evolving mass threshold combined with highly variable mass accretion rates. 

If we imposed such a mass limit at $z=0$ and selected clusters at increasing redshift that also satisfied this cut, very few new members would enter the sample but the number of clusters included would decline rapidly with increasing redshift. By $z=1$ no clusters would be left as at this time the largest object is less massive than the $100^{th}$ biggest object at $z=0$. Alternatively, we could impose a mass limit at $z=1$, say a mass of $3 \times 10^{14}\hmsun$, but in this case we would have thousands of objects in our sample at $z=0$. As \Fig{fig:STcounts} shows, there are expected to be $\sim\!2000$ clusters in the volume with a mass within an order of magnitude of the mass of the largest in the volume. A final alternative is to take the mass limit from the original 324 clusters of \TheThreeHundred, dropping objects from the sample once their mass drops below this limiting mass. There are only 4 clusters more massive than $10^{14.8}\hmsun$ at $z=1$. 

A single mass-completeness limit applied to all redshifts is clearly impractical. Our proposed extension to \TheThreeHundred\ will therefore be complete at all redshifts from $0<z<1$ for the top 100 objects (i.e. with an evolving limiting mass) and the resulting superset will probe a wider range of mass accretion histories and final cluster masses. To reach this requires the addition of only 68 additional clusters to our existing sample of 324 resimulated clusters.  \footnote{In constructing these catalogues it is also important to take into account that two halos in the top 100 at earlier times can undergo a major merger and form a single object at later times and so the number count is not one-to-one conserved. In these cases, both haloes do not need to be resimulated because they form the same final object and all the progenitor particles from both are automatically included in our original Lagrangian region. Some halos are also very close at the final time. We have chosen to only resimulate those that are further than $2\hmpc$ apart at $z=0$}

\section{Summary and Conclusion}\label{sec:conclusion}

In this paper we tested the assumption that the largest objects found within a galaxy cluster sample remain the largest objects at either earlier or later times. We found that this proposition of both theorists and observers is not correct and that a significant number of objects both enter and leave such a sample. 

We find that some haloes experience rapid short term growth, typically via a major merger, and that such an event can significantly raise their rank in the mass hierarchy. Conversely, as these events are stochastic and rare, there can be long periods when a cluster's mass growth appears to stagnate. The rate of any particular cluster's growth can vary a lot about the general mean growth of the population as a whole.

Due to the effective doubling of mass that occurs during a major merger, such an event can propel an object out of the pack to become one of the most massive structures at the given time. This jump is transient however, and the chances of this particular object remaining so outstanding is perhaps unlikely. The object gets overtaken by other examples. This leapfrogging of the lead leads to churn in the most extreme clusters and also produces a tendency for a high fraction of recent mergers amongst the sample of the most massive objects. 

Clusters exhibiting ongoing mergers are dynamically unrelaxed. These objects tend to have recently accreted significant mass and to have jumped up the mass ranking. Around 77\% of the clusters are considered unrelaxed at $z=0$ using the definition in \citet{haggarDS} - this can lead to underestimating the mass by up to 20\%. Conversely, dynamically relaxed objects have been quietly accreting mass more slowly. These objects have not recently rapidly changed position in the mass rankings. 

By tracing the largest objects at all times back to $z=1$ we have constructed an extension to the cluster sample previously used by \TheThreeHundred\ that contains all the largest 100 clusters at every redshift throughout this period. This mass-complete sample of 268 clusters is intended to provide a better match to the evolution of 
observational quantities in large clusters. 

From this work we can draw the following conclusions:
\begin{itemize}
\item{Due to the high churn within the rank order of the largest objects there is limited direct correspondence between the rank order of the large objects at $z=1$ and large objects at $z=0$. While the largest objects at $z=1$ generally remain massive structures they typically lose their outstanding position over time. Less than half of the 100 largest structures in the volume at $z=1$ remain so at $z=0$. The commonly held belief that this correspondence is essentially one-to-one is not the case in practice.}
\item{Selecting a mass-limited cluster sample at a fixed redshift introduces bias. Close to the selection epoch such a sample is likely to contain an enhanced fraction of disturbed systems with a higher than expected fraction of recent or ongoing mergers. The evolutionary histories of these objects will not span all possible histories, being biased towards those with recent rapid mass growth. These are late-forming systems. Early formers, especially if followed by a period of quiescence will likely be underrepresented.}
\item{Selection bias may also extend to environmental bias, although we have not tested this explicitly here. Mass accretion after any initial phase is likely related to the wider environment of the cluster and will vary depending upon the local matter density. Early forming objects in globally underdense regions will therefore likely be depleted in a $z=0$ selected mass limited sample.}
\item{Care should also be taken when analysing a mass complete sample that better mimics an observational selection strategy. Such a sample will also contain a large fraction of mergers, because by construction we are only selecting the largest objects at all times. This is not the same as selecting a random sample of objects from further down the mass function.}
\item{The more unrelaxed a cluster is at $z=0$ the more recently it is likely to have entered the sample.}
\item{By construction, the mass evolution of a unchanging (i.e. fixed) set of objects taken from the extreme of the mass function at a single time will be very different from the mass evolution of a mass complete sample. For an unchanging set of objects the mass growth is more rapid over the period prior to selection (by $\sim (1+z)^{-0.5}$) and slower in the period post selection (by a similar factor) shown visually in \Fig{fig:RSevolve}. Any study that involves mass growth should take this into account.}
\item{Due to the high likelihood of the most massive objects having recently undergone or being in the process of a major merger, finding a rare, large structure should not be proposed as a challenge to the entirety of existing cosmological theory. Such a structure is highly likely to be dynamically unrelaxed and estimating its mass via traditional means is extremely uncertain and likely to be strongly biased to the high side as seen in \Fig{fig:HaloMerger} and {\ref{fig:dstate}}.}
\end{itemize}

In conclusion, caution should be exercised when using a small mass-selected sample of the most extreme objects. Such a sample will be subject to change with respect to objects from further down the mass function in terms of recent mass growth, future mass growth and current environmental state. As \Fig{fig:RSevolve} illustrates, the largest halo at any epoch can only ever stay at the top or slip down the rank, other haloes from further down the mass function are free to adjust their ranking position in either direction.

There is no guarantee that objects that are currently the most massive were also the most massive in the past or will remain so in the future. 

\section*{Acknowledgements}
We acknowledge access to the theoretically modelled galaxy cluster data via The Three Hundred (https://the300-project.org) collaboration. The simulations used in this paper have been performed in the MareNostrum Supercomputer at the Barcelona Super-computing Center, thanks to CPU time granted by the Red Espa\^{n}ola de Supercomputaci\'{o}n. As part of The Three Hundred project, this work has received financial support from the European Union’s Horizon 2020 Research and Innovation programme under the Marie Sk\l{}odowska-Curie grant agreement number 734374, the LACEGAL project. 
AK is supported by the Ministerio de Ciencia e Innovaci\'{o}n (MICINN) under research grant PID2021-122603NB-C21 and further thanks This Mortal Coil for filigree \& shadow.
WC is supported by Atracci\'{o}n de Talento Contract no. 2020-T1/TIC19882 granted by the Comunidad de Madrid and by the Consolidación Investigadora grant no. CNS2024-154838 granted by  the Agencia Estatal de Investigación (AEI)  in Spain. He also thanks the Ministerio de Ciencia e Innovaci\'{o}n (Spain) for financial support under Project grant PID2021-122603NB-C21, ERC: HORIZON-TMA-MSCA-SE for supporting the LACEGAL-III Latin American Chinese European Galaxy Formation Network) project with grant number 101086388, and  the science research grants from the China Manned Space Project.

The authors contributed to this paper in the following way. JO, FP \& AK wrote the paper and carried out most of the analysis. GY produced the initial conditions and ran the simulations together with WC. RH supplied the dynamical state information. ACS provided the major merger catalogue. MG \& UK provided suggestions and guidance throughout the project.

We are grateful for the helpful comments provided by the referee.
\section*{Data Availability}
Data will be shared on request to the corresponding author. Some of the results shown in this work use data from TheThreeHundred galaxy clusters sample. These data are available on request following the guidelines of TheThreeHundred collaboration, at https://www.the300-project.org. 

\bibliography{mn-jour,WWIB}
\bibliographystyle{mn2e} 
\label{sec:Bibliography}

\bsp
\label{lastpage}

\end{document}